A High Pressure Neutron Study of Colossal Magnetoresistant NdMnAsO$_{0.95}$F$_{0.05}$

E. J. Wildman [1], M. G. Tucker [2] and A. C. Mclaughlin* [1]

[1] The Chemistry Department, University of Aberdeen, Meston Walk, Aberdeen, AB24 3UE, Scotland.

[2] ISIS Facility, Rutherford Appleton Laboratory, Harwell, Didcot OX11 0DE, UK.

*a.c.mclaughlin@abdn.ac.uk

**Abstract**

A high pressure neutron diffraction study of the oxypnictide NdMnAsO$_{0.95}$F$_{0.05}$ has been performed at temperatures of 290 K – 383 K and pressures up to 8.59 GPa. The results demonstrate that the antiferromagnetic order of the Mn spins is robust to pressures of up to 8.59 GPa. T$_N$ is enhanced from 360 K to 383 K upon applying an external pressure of 4.97 GPa, a rate of 4.63 K /GPa. NdMnAsO$_{0.95}$F$_{0.05}$ is shown to violate Bloch's rule which would suggest that NdMnAsO$_{0.95}$F$_{0.05}$ is on the verge of a localised to itinerant transition. There is no evidence of a structural transition but applied pressure tends to result in more regular As-Mn-As and Nd-O-Nd tetrahedra. The unit cell is significantly more compressible along the *c*-axis than the *a*-axis, as the inter-layer coupling is weaker than the intrinsic bonds contained within NdO and MnAs slabs.

**Introduction**

Due to the recent discovery of high temperature superconductivity at 26 K in electron doped LaFeAsO [1] there has been much active research into oxypnictide materials. These layered 1111-type pnictides form with the primitive tetragonal ZrCuSiAs structure of space group *P4/nmm*. Enhancing the superconducting transition temperature of this family of compounds has been achieved by using high pressure synthesis to create oxygen vacancies [2] and also by carrying out substitutions on the rare earth site, with the maximum $T_c$ of 56.3 K currently held by $Gd_{1-x}Th_xFeAsO$. [3] Variable pressure X-ray diffraction results carried out at room temperature on the series $LaFeAsO_{1-x}F_x$ [4, 5] at pressures ≤ 10 GPa revealed no structural transition and tetragonal symmetry was preserved over the range studied. Anisotropic compression occurs as the axial compressibility of the *c* axis ($\kappa_c$) is almost twice that observed along *a* ($\kappa_a$). The crystallographic direction normal to the FeAs layer is more effected by pressure than that parallel to the FeAs layer, and both structural parameters decrease monotonically so that the c/a ratio decreases linearly with pressure. It has been established that the anisotropic compression changes the charge distributions within the $[FeAs]^-$ and $[LnO]^+$ layers, allowing the electronic properties to be easily manipulated upon applying external pressure. For $LnFeAsO_{1-y}$ (*Ln* = La, Nd, Pr, Sm) the value of $T_c$ increases as the bond angles within the $FeAs_4$ tetrahedron tend towards optimised values, and the maximum $T_c$ is achieved when a regular tetrahedron is adopted with angles of 109.47°. [6]

It is also possible to synthesise pnictides and oxypnictides so that other transition metals, such as Mn, Co or Ni replace Fe. [7, 8, 9] Manganese pnictides have received particular attention due to their interesting magnetic and electronic properties. $BaMn_2As_2$ is a G-type antiferromagnetic insulator, within which the Mn spins order antiferromagnetically in a 'checker-board' fashion aligned along *c*. [10] Pressure induced metallization occurs at 36 K when ~ 4.5 GPa is applied to the material and above 5.8 GPa the compound is metallic over the entire temperature range. [11] X-ray diffraction studies carried out as a function of pressure revealed an anomaly in the unit cell volume at ~ 5 GPa which was not accompanied by a change in crystal structure, indicating an electronic transition that is consistent with the resistivity results.

Colossal magnetoresistance has recently been reported in the 1111-type manganese pnictide series for $NdMnAsO_{1-x}F_x$, with a maximum CMR of -95% achieved at 3 K when x = 0.05. [12] Magnetoresistance (MR) is defined as the change of electrical resistivity $\rho$ in an applied magnetic field $H$, so that MR= $(\rho(H)-\rho(0))/\rho(0)$, where *ρ(H)* and *ρ(0)* are the resistivities in an applied field and zero field respectively. Magnetoresistant materials are important for magnetic memory device applications. Variable temperature neutron diffraction measurements on $NdMnAsO_{0.95}F_{0.05}$ shows the same magnetic structure to that originally reported for the parent compound NdMnAsO (Fig. 1).[13, 14] AFM ordering arises below 356(2) K ($T_N$ (Mn)), with Mn moments aligned parallel to the *c* axis. At 23 K ($T_N$(Nd)) the $Nd^{3+}$ spins order with moments parallel to the basal plane, which results in a spin re-orientation ($T_{SR}$) of Mn spins as they rotate from their previous alignment along *c* to along *a*. At the same time an Efros Schlovskii transition is observed which suggests that the reorientation of Mn spins into the basal plane at 20 K results in enhanced Coulomb correlations between localized electrons, resulting in a much higher resistivity below $T_{SR}$. [12] A variable field neutron diffraction study has shown that upon applying a magnetic field there is a second order phase transition from antiferromagnetic to paramagnetic order of both the Nd and Mn spins. [12] The CMR then arises as a result of a second order phase transition from an insulating antiferromagnet to a semiconducting paramagnet upon applying a magnetic field, so that the electron correlations are diminished in field.

It has been suggested that an antiferromagnetic instability is present in NdMnAsO$_{0.95}$F$_{0.05}$.[12] LaMnPO has very similar magnetic properties to NdMnAsO$_{0.95}$F$_{0.05}$ (T$_N$ – 375 K).[15] A high pressure resistivity study on LaMnPO has revealed that an insulator-to-metal transition occurs at 20 GPa along with the complete suppression of long-ranged AFM order at higher pressures of 32 GPa. In this study we report the effect of pressure on the structure and antiferromagnetic ordering transition of NdMnAsO$_{0.95}$F$_{0.05}$.

**Methods**

A 1g polycrystalline sample of NdMnAsO$_{0.95}$F$_{0.05}$ was synthesised *via* a two-step solid-state reaction. Initially, the NdAs precursor was obtained by the reaction of Nd pieces (Aldrich 99.9%) and As (Alfa Aesar 99.999%) at 900$^o$C for 24h in an evacuated, sealed quartz tube. The resulting precursor was then reacted with stoichiometric amounts of MnO$_2$, Mn and MnF$_2$ (Aldrich 99.99%), all powders were ground in an inert atmosphere and pressed into pellets of 10mm diameter. The pellets were placed into a Ta crucible and sintered at 1150$^o$C for 48h, again in a quartz tube sealed under vacuum.

Powder X-ray diffraction patterns of NdMnAsO$_{0.95}$F$_{0.05}$ were collected using a Bruker D8 Advance diffractometer with twin Gobel mirrors and Cu K$\alpha$ radiation. Data were collected at room temperature over the range 10 < 2$\theta$ < 100$^o$, with a step size of 0.02$^o$, and could be indexed on a tetragonal unit cell of space group *P4/nmm*, characteristic of the ZrCuSiAs structure type. X-ray diffraction patterns demonstrated that the material was of high purity. Powder neutron diffraction patterns were recorded on the high intensity diffractometer D20 at the ILL, Grenoble with a wavelength of 2.4188 Å. A 1 g sample of NdMnAsO$_{0.95}$F$_{0.05}$ was inserted into an 8 mm vanadium can and data were recorded at selected temperatures between 350 K and 400 K with a collection time of 10 minutes per temperature.

High pressure time-of-flight neutron diffraction patterns were recorded using the instrument PEARL at the ISIS facility, U.K. The sample was loaded into an encapsulated TiZr gasket [16] with 4:1 methanol–ethanol used as the pressure transmitting medium. A small pellet of lead was placed into the cell with the sample for use as a pressure calibrant. [17] The sample was then loaded into the Paris–Edinburgh cell. [18] The cell was used in transverse geometry giving access to scattering angles in the range 81.2° < 2θ < 98.8°. Data were recorded at room temperature using pressures up to ~ 8.6 GPa with a collection time of ~ 4 hour per pressure. Data sets were also recorded at temperatures between 360 K - 383 K using pressures of up to 4.97 GPa. The lead equation of state (EOS) [19] used to calculate the pressure was a Birch–Murnaghan equation of the form:

$$\frac{V}{V_0} = \left(1 + \frac{B'P}{B_0}\right)^{-1/B'}$$

where $V_0$ is the unit cell volume at zero pressure, $V$ the unit cell volume at pressure $P$, $B_0$ the zero pressure bulk modulus and $B'$ is the pressure derivative of the ambient bulk modulus. For Pb, the values were taken to be, $B_0$ = 42 GPa, $B'$ = 5. A wavelength-dependent attenuation correction [20] was applied to account for the different sample environment materials before the data were analysed.

The pressure dependency of the structure was obtained by Rietveld refinement [21] of the neutron data using the GSAS program. [22]

**Results and Discussion**

Figure 1 shows the magnetic structure and the 290 K neutron diffraction pattern and Rietveld fit obtained for NdMnAsO$_{0.95}$F$_{0.05}$ at a pressure of 3.65 GPa. There is no evidence of a phase transition up to 8.59 GPa, which is common for other 1111 oxypnictides of this structure type (Fig 2). [23] A tetragonal *P4/nmm* unit cell was observed over the entire pressure range with fully occupied cation and anion sites. The Pr, Mn and As occupancies refined to within ± 1 % of the full occupancy and were fixed at 1.0. The O and F occupancies were fixed at 0.95 and 0.05 respectively. The corresponding fit parameters, refined lattice constants, bond lengths and angles at each pressure are given in Table 1. The (101) and (100) magnetic peaks can be observed, alongside a magnetic contribution to the (102) and (103) structural peaks (Figures 1 and 2), with no change in magnetic structure upon increasing the pressure. The Mn moment refines to 1.92(7) $\mu_B$ and 1.93(7) $\mu_B$ at $P$ = 0 GPa and 8.59 GPa respectively (Table 1). There is no evidence of a spin reorientation upon increasing the pressure so that the Mn spins remain aligned parallel to *c* for all *P*.

The cell parameters decrease upon increasing pressure and the variation in unit cell volume ($V/V_o$) with pressure is shown in Figure 3. The axial compressibility of the lattice parameters, $k$, were calculated using linear fits of the normalised lattice constants, $a/a_o$ and $c/c_o$ (where for example $k_a$ = -1$a$(d$a$/d$P$) etc.). The obtained values of $k_a$ = 3 x 10$^{-3}$ GPa$^{-1}$ and $k_c$ = 5.2 x 10$^{-3}$ GPa$^{-1}$ indicate that the unit cell is significantly more compressible along the *c*-axis than the *a*-axis, as the inter-layer coupling is weaker than the intrinsic bonds contained within [NdO]$^+$ and [MnAs]$^-$ slabs. The bulk modulus, $B_0$, of NdMnAsO$_{0.95}$F$_{0.05}$ was calculated using a single parameter Birch-Murnaghan fit (shown as the red line in Fig. 3), with the volume of the unit cell at ambient pressure fixed at the refined value of 145.81(1) Å$^3$. The resultant fits gave values of $B_0$ = 74(1) GPa and $B_0'$ = 3.4(5), which are similar to the values extracted from pressure studies of other 1111-type pnictides (for example, $B_0$ = 70 GPa for LaFeAsO$_{0.95}$F$_{0.05}$ [4]).

The Nd-O and Mn-As bond lengths also decrease upon applying pressure, *P* (Table 1). The change in thickness of the respective layers is however interesting. The MnAs layers become more compressed as pressure is applied, shrinking by ~ 3.5 % upon increasing $P$ from 0 – 8.59 GPa. In contrast the NdO layer expands by ~2.2 % (Figure 4). The results demonstrate that in NdMnAsO$_{0.95}$F$_{0.05}$ the interlayer spacing undergoes a reduction of ~ 9.6 % upon increasing $P$ from 0 - 8.9 GPa (Table 1). The atomic positions of Mn and (O, F) are constrained by symmetry (Wyckoff position 2b and 2a, respectively), whereas both Nd and As are located at 2c and hence their z atomic position can be affected by external pressure. Upon increasing the pressure from 0 – 8.59 GPa the $z$ co-ordinate of Nd increases from 0.1295(3) to 0.1385(5). At the same time the z co-ordinate of As changes from 0.6734(5) to 0.6750(6) (Table 1) and as a result of these combined effects the Nd-As bond length shrinks by 4.4 %. The effect of external pressure therefore is to bring the [MnAs]$^-$ and [NdO]$^+$ layers together. The variation of $\alpha1$ and $\alpha2$ Nd-O/F-Nd bond angles with applied pressure are shown in Figure 5 and Table 1. The Nd-O/F-Nd bond angles change more towards an ideal tetrahedron, as $\alpha1$ decreases from 120.7(1)° to 118.3(1)°, while $\alpha2$ increases from 104.16(6)° to 105.25(7)° upon increasing $P$ from 0 – 8.59 GPa. The consequence of the change in the Nd z atomic position and changes in the $\alpha1$ and $\alpha2$ Nd-O/F-Nd bond angles is therefore that the NdO layer expands very slightly with pressure.

Figure 6 shows the variation of $\alpha1$ and $\alpha2$ As-Mn-As angles with applied pressure. In contrast to the change in Nd-O/F-Nd bond angles described above, a much smaller variation of the As-Mn-As bond angles with $P$ is observed. Over the pressure range studied for NdMnAsO$_{0.95}$F$_{0.05}$, the As-Mn-As bond angles tend very gradually towards an ideal tetrahedron, as $\alpha1$ decreases from 111.55(8)° to 111.3(1)°, while $\alpha2$ increases from 105.4(1)° to 105.9(1)° as $P$ increases from 0 GPa to 8.59 GPa. The change in As-Mn-As bond angle with pressure seems to be greater above $P$ = 5.51 GPa and much higher pressures would be required in order to observe if the optimum tetrahedron can be obtained.

Variable temperature, variable pressure measurements were also carried out using the Paris–Edinburgh cell. The sample was heated to temperatures above $T_N$ = 360 K where there was no apparent magnetic diffraction. The pressure was then increased at each temperature until the (101) and (100) magnetic diffraction peaks reappeared (Fig. 7, inset b). $T_N$ (at a specific temperature above $T_N$ = 360 K at P = 0) is defined as the pressure required in order to re-establish magnetic diffraction. Magnetic diffraction was re-established at 370 K (2.08 GPa), 380 K (4.23 GPa) and 383 K (4.97 GPa). The results show that it is possible to increase $T_N$ from 360 K to 383 K upon applying an external pressure of 4.97 GPa, a rate of 4.63 K /GPa (Fig. 7).

It is likely that the anisotropic compression of the NdMnAsO$_{0.95}$F$_{0.05}$ structure upon application of external pressure results in changes in charge distributions within the [MnAs]$^-$ and [NdO]$^+$ slabs [23]. However it has been shown that neither hole doping or electron doping results in an increase in $T_{Mn}$ [24, 12], therefore it is highly unlikely that the enhanced $T_{Mn}$ upon increasing external pressure is a result of a change in the charge distribution within the [MnAs]$^-$ layer. Instead it is more likely to be a structural effect. Table 2 shows the variation of the structural parameters with temperature and pressure for NdMnAsO$_{0.95}$F$_{0.05}$. It is clear that the Mn-As bond length decreases upon increasing both temperature and pressure from 290 K, 0 GPa to 383 K, 4.68 GPa. At the same time the interlayer spacing also decreases from 1.754(5) Å to 1.622(9) Å. There is no real change in the $\alpha 1$ or $\alpha 2$ As-Mn-As bond angles.

In localised antiferromagnets, the Heisenberg exchange interaction, $J_H$, has been shown to increase with applied pressure as a result of better orbital overlap as the cation-anion bond length decreases. Numerous antiferromagnetic insulators have been shown to obey Bloch's rule, where $\alpha$ = $d$log($T_N$)/$d$log($V$) ~ -3.3 where $V$ is the cell volume. [25] A theoretical rationalisation of Bloch's rule comes from calculations of the variation of the overlap integral with the cation-anion bond length. Neutron scattering studies of NdMnAsO$_{0.95}$F$_{0.05}$ recorded on diffractometer D20 at the ILL demonstrate that above T$_{Mn}$ there is magnetic diffuse scattering, characteristic of short range magnetic order up to 400 K (Fig. 7a, bottom inset). Upon application of pressure, the reduction in both the Mn-As bond length and the interlayer spacing enhances the superexchange between Mn centres, both along Mn-As-Mn and between the planes, which then results in the increase in T$_{Mn}$. The variation of log($T_N$) against log($V$) is shown in the top inset to Figure 7a and a small value of $\alpha$ = -0.95(2) is obtained, violating Bloch's law. The perturbation description for the superexchange spin-spin interaction should break down on the approach to crossover from localized to itinerant electronic behaviour of a Mott-Hubbard insulator. The small value of $\alpha$ = -0.95(2) would suggest that NdMnAsO$_{0.95}$F$_{0.05}$ is on the verge of a localised to itinerant transition so that a purely localised Heisenberg description of the magnetic exchange is not applicable. This is further corroborated by variable field resistivity and neutron diffraction measurements which have shown that competing electronic phases are present in NdMnAsO$_{0.95}$F$_{0.05}$.[12]

CMR is observed in NdMnAsO$_{0.95}$F$_{0.05}$ below the spin reorientation transition, T$_{SR}$, of the Mn spins which is precipitated by the antiferromagnetic ordering of the Nd$^{3+}$ spins at T$_{Nd}$.[12] It is possible that T$_{Nd}$ will also increase with applied pressure which may result in CMR observed at higher temperatures. High pressure neutron diffraction and magnetoresistance measurements at low temperature are warranted to investigate this further.

LaMnPO also has an AFM ordered state, with the same antiferromagnetic structure as NdMnAsO$_{0.95}$F$_{0.05}$ and a comparable T$_N$(Mn) of 375 K.[15] Hence it is worthwhile comparing the effects of pressure on T$_N$(Mn). Pressure studies on this insulating material revealed a contrasting suppression of T$_N$ with increasing pressure (T$_N$ decreases from 375 K to 290 K upon applying a pressure of 7.3 GPa). An induced crossover to a mixed state, in which insulating and metallic states coexist, is observed upon applying a pressure of 20 GPa followed by the collapse of long range AFM order at ~32 GPa. The reduction in $T_N$ with pressure is therefore a result of increasing electron delocalisation upon increasing pressure and by 32 GPa the localised/moment bearing electrons are fully delocalised. A high pressure X-ray diffraction study of the same compound shows that the tetragonal structure is stable up to 16.4 GPa. [26] At higher pressures an orthorhombic structure is observed followed by a collapsed orthorhombic state at 31 GPa.

LaMnPO is reported to be less electronically stable than its substantial gap and ordered moment suggest, and, at ambient pressure, is close to an electron delocalisation transition (EDT) that is driven by the nucleation of states with energies within the correlation gap [15, 26]. The electronic structure of $NdMnAsO_{0.95}F_{0.05}$ appears to be more robust to electronic delocalisation upon applying external pressure, so that at modest pressures the shorter Mn-As bonds and interlayer spacing results in enhanced superexchange between Mn centres and higher $T_N$ upon increasing $P$. However the small value of $\alpha$ = -0.95(2) obtained from the fit to the Bloch equation shows that Bloch's rule is violated in $NdMnAsO_{0.95}F_{0.05}$ and that $NdMnAsO_{0.95}F_{0.05}$ is also on the verge of a localised to itinerant transition. Presumably, eventually in high enough pressures, an electron delocalisation transition will occur and $T_N$ will decrease. Further neutron diffraction experiments with higher pressures will be necessary to see if there is an eventual collapse in the antiferromagnetic order and/or structural phase change with increasing pressure. Electronic structure calculations and variable pressure resistivity measurements are also warranted to explore this further.

In summary we show that the antiferromagnetic order of the Mn oxyarsenide $NdMnAsO_{0.95}F_{0.05}$ is robust to pressures of up to 8.9 GPa and that $T_N$ is enhanced with applied pressure at a rate of 4.63 K /GPa. This is in contrast to the oxyphosphide LaMnPO, where $T_N$ decreases in modest pressure. There is also no evidence of a crystallographic phase change or change in magnetic structure upon increasing the pressure up to 8.9 GPa.

**Acknowledgements**

We acknowledge the UK EPSRC for financial support (Grant EP/L002493/1) and STFC-GB for provision of beamtime at ISIS and ILL.

Table 1: Refined cell parameters, agreement factors, atomic parameters and selected bond lengths and angles for NdMnAsO$_{0.95}$F$_{0.5}$ from Rietveld fits against neutron diffraction data at various pressures. Nd and As are at 2c (¼, ¼, z), Mn at 2b (¾, ¼, ½) and O,F at 2a (¾, ¼, 0).

| Atom | Occupancy | | Pressure (GPa) | | | | | | |
|---|---|---|---|---|---|---|---|---|---|
| | | | 0 | 0.46 | 2.03 | 3.65 | 5.51 | 7.55 | 8.59 |
| Nd | 1.00 | z | 0.1295(3) | 0.1304(2) | 0.1319(2) | 0.1338(3) | 0.1349(3) | 0.1371(3) | 0.1385(5) |
| | | $U_{iso}$ (Å$^2$) | 0.007(1) | 0.007(1) | 0.0016(9) | 0.002(1) | 0.001(1) | 0.001(1) | 0.004(1) |
| Mn | 1.00 | $U_{iso}$ (Å$^2$) | 0.003(1) | 0.009(1) | 0.010(1) | 0.005(1) | 0.001(1) | 0.005(1) | 0.003(1) |
| As | 1.00 | z | 0.6734(5) | 0.6734(3) | 0.6740(4) | 0.6744(4) | 0.6751(4) | 0.6757(4) | 0.6750(6) |
| | | $U_{iso}$ (Å$^2$) | 0.011(1) | 0.0114(9) | 0.010(1) | 0.008(1) | 0.006(1) | 0.010(1) | 0.009(1) |
| O/F | 0.95/0.05 | $U_{iso}$ (Å$^2$) | 0.009(1) | 0.011(1) | 0.008(1) | 0.005(1) | 0.004(1) | 0.008(1) | 0.009(1) |
| | | a (Å) | 4.04840(6) | 4.04148(4) | 4.01940(5) | 3.99869(6) | 3.97672(8) | 3.95391(7) | 3.9445(2) |
| | | c (Å) | 8.8965(6) | 8.8728(4) | 8.7913(5) | 8.7117(7) | 8.6314(8) | 8.5301(7) | 8.5032(8) |
| | | $\chi^2$ (%) | 0.539 | 0.499 | 0.453 | 0.500 | 0.484 | 0.524 | 0.467 |

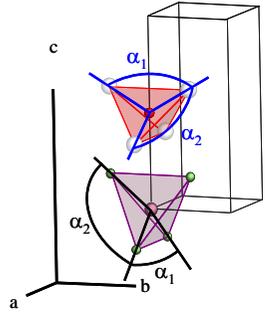

| | | | | | | | |
|---|---|---|---|---|---|---|---|
| $R_{WP}$ (%) | 4.14 | 3.09 | 3.25 | 3.42 | 3.51 | 3.07 | 4.49 |
| $R_P$ (%) | 5.25 | 3.84 | 3.96 | 4.30 | 4.27 | 3.75 | 5.34 |
| Nd-O/F (Å) | 2.329(1) | 2.329(1) | 2.320(1) | 2.314(1) | 2.304(1) | 2.297(1) | 2.297(1) |
| Mn-As (Å) | 2.545(3) | 2.540(2) | 2.526(2) | 2.511(2) | 2.498(2) | 2.481(2) | 2.471(3) |
| Mn-Mn (Å) | 2.86265(4) | 2.85776(3) | 2.84215(3) | 2.82750(4) | 2.81197(5) | 2.79584(5) | 2.7892(1) |
| Nd-As (Å) | 3.357(3) | 3.346(2) | 3.315(2) | 3.285(2) | 3.255(2) | 3.220(2) | 3.208(3) |
| $\alpha_1$ Nd-O/F-Nd (°) | 120.7(1) | 120.41(9) | 120.03(9) | 119.5(1) | 119.3(1) | 118.8(1) | 118.3(1) |
| $\alpha_2$ Nd-O/F-Nd (°) | 104.16(6) | 104.29(4) | 104.46(4) | 104.69(4) | 104.79(5) | 105.03(4) | 105.25(7) |
| $\alpha_1$ As-Mn-As (°) | 105.4(1) | 105.4(1) | 105.4(1) | 105.6(1) | 105.5(1) | 105.7(1) | 105.9(1) |
| $\alpha_2$ As-Mn-As (°) | 111.55(8) | 111.52(6) | 111.52(6) | 111.47(7) | 111.48(7) | 111.41(6) | 111.3(1) |
| MnAs Layer (Å) | 3.085(5) | 3.077(4) | 3.059(4) | 3.039(6) | 3.023(7) | 2.997(6) | 2.976(7) |
| Nd(O/F) Layer (Å) | 2.304(5) | 2.314(4) | 2.319(5) | 2.331(6) | 2.329(7) | 2.339(6) | 2.355(7) |
| Interlayer spacing (Å) | 1.754(5) | 1.741(4) | 1.707(4) | 1.671(5) | 1.640(5) | 1.597(6) | 1.586(7) |
| Mn moment ($\mu_B$) | 1.92(7) | 1.87(6) | 1.89(5) | 1.99(5) | 1.98(5) | 1.99(5) | 1.93(7) |

Table 2: Refined cell parameters, agreement factors, atomic parameters and selected bond lengths and angles for NdMnAsO$_{0.95}$F$_{0.5}$ from Rietveld fits against neutron diffraction data at various temperatures and pressures. Nd and As are at 2c (¼, ¼, z), Mn at 2b (¾, ¼, ½) and O,F at 2a (¾, ¼, 0).

| Atom | Occupancy | | Temperature (K) / Pressure (GPa) | | | |
|---|---|---|---|---|---|---|
| | | | 290 / 0 | 370 / 2.08 | 380 / 4.23 | 383 / 4.97 |
| Nd | 1.00 | z | 0.1295(3) | 0.1328(3) | 0.1355(4) | 0.1353(4) |
| | | $U_{iso}$ (Å$^2$) | 0.007(1) | 0.002(1) | 0.010(1) | 0.001(1) |
| Mn | 1.00 | $U_{iso}$ (Å$^2$) | 0.003(1) | 0.015(1) | 0.014(2) | 0.011(2) |
| As | 1.00 | z | 0.6734(5) | 0.6749(4) | 0.6781(6) | 0.6767(6) |
| | | $U_{iso}$ (Å$^2$) | 0.011(1) | 0.0104(9) | 0.014(1) | 0.015(1) |
| O/F | 0.95/0.05 | $U_{iso}$ (Å$^2$) | 0.009(1) | 0.008(1) | 0.008(1) | 0.006(1) |
| | | *a (Å)* | 4.04840(6) | 4.015272(6) | 3.98956(9) | 3.97916(9) |

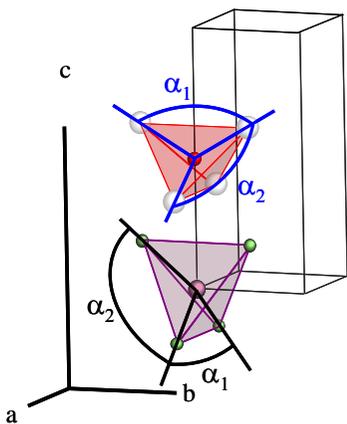

| | | | | |
|---|---|---|---|---|
| $c$ (Å) | 8.8965(6) | 8.7728(6) | 8.654(1) | 8.628(1) |
| $\chi^2$ (%) | 0.539 | 1.037 | 1.149 | 0.916 |
| $R_{WP}$ (%) | 4.14 | 2.67 | 3.77 | 3.50 |
| $R_P$ (%) | 5.25 | 2.84 | 4.09 | 3.84 |
| Nd-O/F (Å) | 2.329(1) | 2.321(1) | 2.314(2) | 2.307(2) |
| Mn-As (Å) | 2.545(3) | 2.527(2) | 2.521(3) | 2.507(3) |
| Mn-Mn (Å) | 2.86265(4) | 2.83923(4) | 2.82105(7) | 2.81369(7) |
| Nd-As (Å) | 3.357(3) | 3.302(2) | 3.250(3) | 3.247(3) |
| $\alpha_1$ Nd-O/F-Nd (°) | 120.7(1) | 119.74(9) | 119.1(2) | 119.2(1) |
| $\alpha_2$ Nd-O/F-Nd (°) | 104.16(6) | 104.59(4) | 104.88(7) | 104.84(6) |
| $\alpha_1$ As-Mn-As (°) | 111.55(8) | 111.63(6) | 112.0(1) | 111.73(9) |
| $\alpha_2$ As-Mn-As (°) | 105.4(1) | 105.2(1) | 104.6(2) | 105.1(2) |
| MnAs Layer | 3.085(5) | 3.068(5) | 3.082(9) | 3.049(9) |
| Nd(O/F) Layer | 2.304(5) | 2.330(5) | 2.345(9) | 2.335(9) |
| Interlayer spacing (Å) | 1.754(5) | 1.687(5) | 1.614(9) | 1.622(9) |

**Figure Captions**

Fig. 1 Rietveld refinement fit to the 290 K, $P$ = 3.65 GPa PEARL neutron powder diffraction pattern of NdMnAsO$_{0.95}$F$_{0.05}$. Tick marks represent reflection positions for NdMnAsO$_{0.95}$F$_{0.05}$ (magnetic structure), ZrO$_2$ (from the ceramic anvil), Al$_2$O$_3$ (from the ceramic anvil), Pb and NdMnAsO$_{0.95}$F$_{0.05}$ from top to bottom respectively. The (101) and (100) magnetic peaks and (102) and (102) magnetic and nuclear peaks are highlighted. The inset shows the magnetic structure at 290 K, where the black spheres represent Mn and the grey spheres represent Nd.

Fig. 2 Neutron diffraction patterns of NdMnAsO$_{0.95}$F$_{0.05}$ recorded at 290 K at pressures between 0 – 8.59 GPa as labelled. The Rietveld fit to the pattern is displayed at each pressure.

Fig. 3 (colour online) Pressure dependence of the normalized unit-cell volume. The line represents the fit to the Birch–Murnaghan equation.

Fig. 4 The variation of the MnAs and NdO layer thickness with applied pressure.

Fig. 5 The dependence of the $\alpha$1 and $\alpha$2 Nd-O-Nd bond angles with pressure which show a clear change to a more regular tetrahedron upon applying pressure.

Fig. 6 The pressure variation of the $\alpha$1 and $\alpha$2 As-Mn-As bond angles which show that there is a gradual change to a more regular MnAs$_4$ tetrahedron upon increasing pressure.

Fig. 7 (a) The variation of $T_N$ with applied pressure showing that $T_N$ increases at a rate of 4.63 K /GPa. The top inset shows the variation of log($T_N$) against log($V$). The bottom inset shows the ambient pressure, 380 K neutron diffraction pattern recorded on diffractometer D20 at the ILL. The diffuse scattering observed above $T_N$ is indicated. (b) The 380 K, 4.23 GPa neutron diffraction pattern, recorded on the instrument PEARL at the ISIS facility, UK. Tick marks represent reflection positions for NdMnAsO$_{0.95}$F$_{0.05}$ (magnetic structure), ZrO$_2$ (from the ceramic anvil), $Al_2O_3$ (from the ceramic anvil), Pb and $NdMnAsO_{0.95}F_{0.05}$ from top to bottom respectively. The (101) magnetic diffraction peak is indicated.

Figure 1

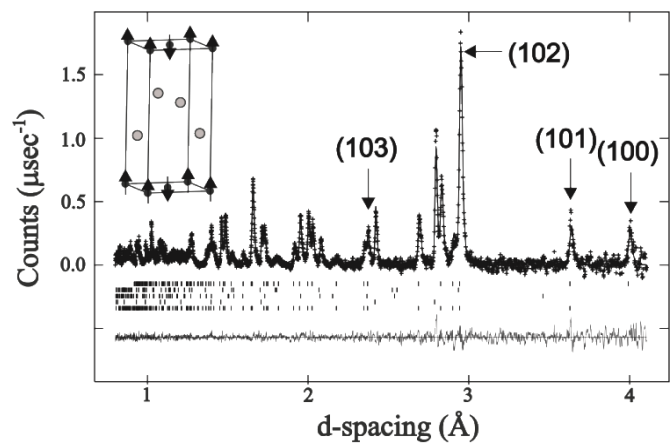

Fig. 2

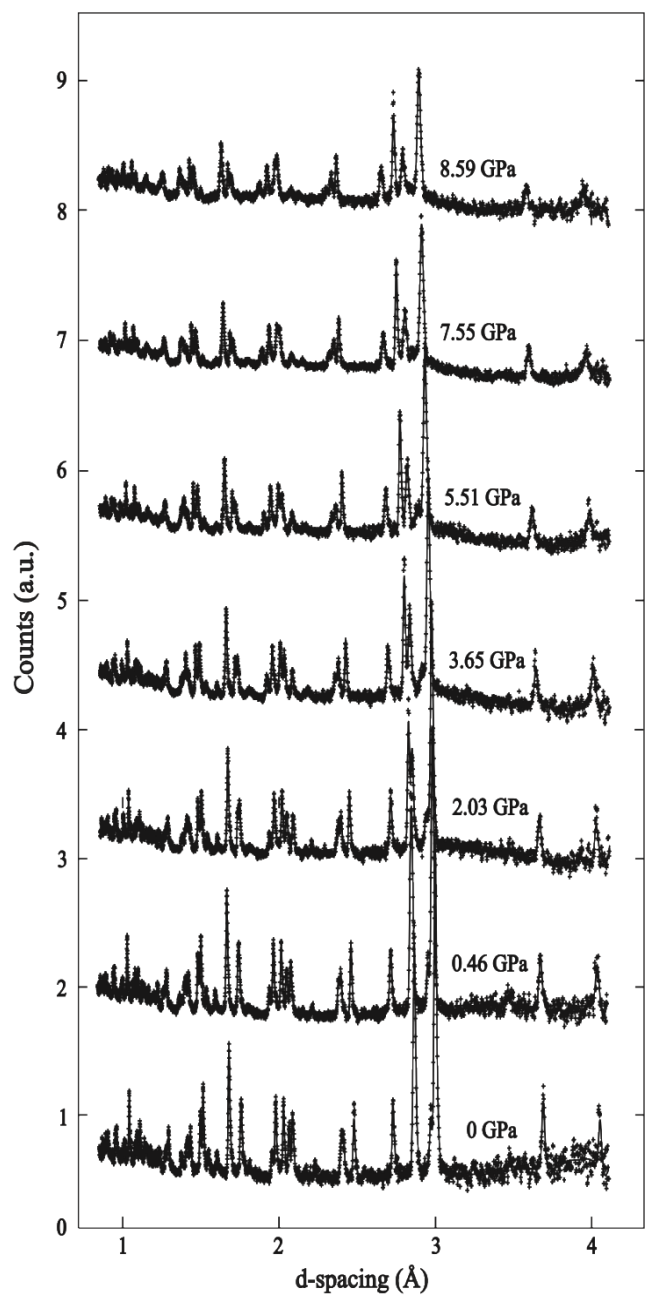

Fig. 3

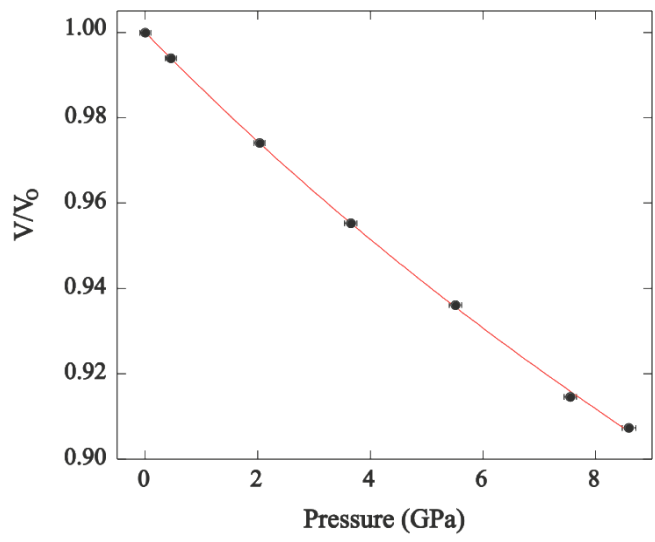

Fig. 4

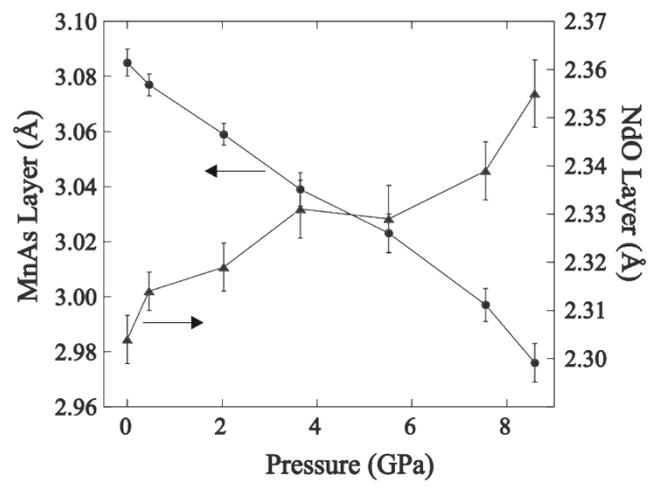

Fig. 5

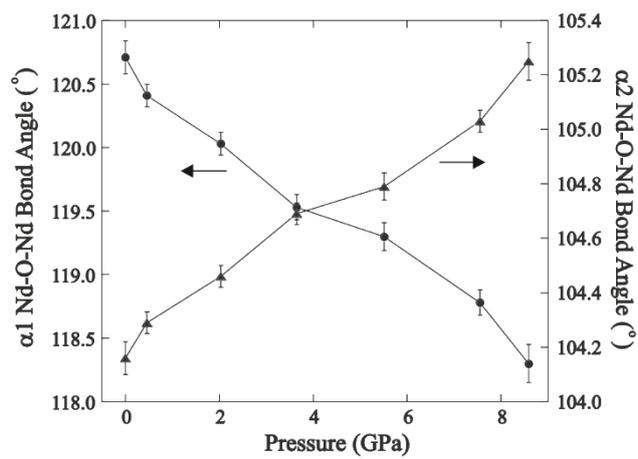

Fig. 6

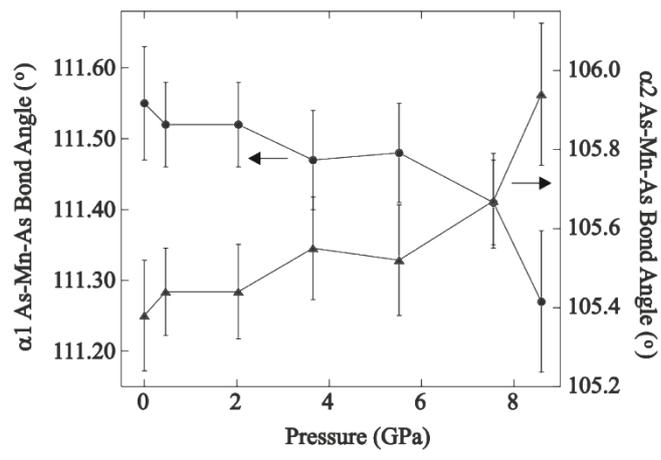

Fig. 7

a
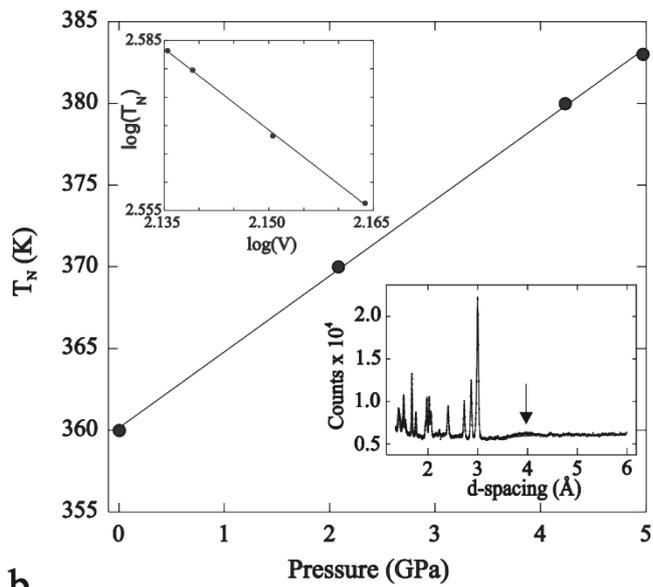

b
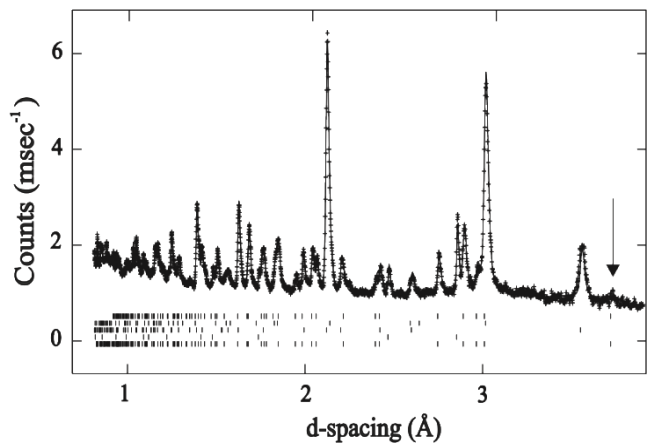